\documentclass[aps,prd,twocolumn,amsmath,superscriptaddress,amssymb,showpacs,floatfix,nofootinbib,longbibliography]{revtex4-1}
\pdfoutput=1
\usepackage{graphicx}
\usepackage{bm}
\usepackage{times}
\usepackage{slashed}
\usepackage{color}
\usepackage{aas_macros}
\usepackage{slashed}
\usepackage{lipsum}
\usepackage{subfigure}
\usepackage{multirow}
\usepackage{amsmath}
\usepackage{array} 
\usepackage{varwidth} 
\begin{document}

\title{Anisotropic diffusion cannot explain TeV halo observations}

\author{Pedro De La Torre Luque}
\email{pedro.delatorreluque@fysik.su.se, ORCID: 0000-0002-4150-2539}
\affiliation{Stockholm University and The Oskar Klein Centre for Cosmoparticle Physics,  Alba Nova, 10691 Stockholm, Sweden}

\author{Ottavio Fornieri}
\email{ottavio.fornieri@gssi.it, ORCID: 0000-0002-6095-9876}
\affiliation{Gran Sasso Science Institute, Viale Francesco Crispi 7, 67100 L’Aquila, Italy}
\affiliation{INFN Laboratori Nazionali del Gran Sasso, Via G. Acitelli 22, 67100 Assergi (AQ), Italy}

\author{Tim Linden}
\email{linden@fysik.su.se, ORCID: 0000-0001-9888-0971}
\affiliation{Stockholm University and The Oskar Klein Centre for Cosmoparticle Physics,  Alba Nova, 10691 Stockholm, Sweden}

\begin{abstract}
TeV halos are regions of enhanced photon emissivity surrounding pulsars. While multiple sources have been discovered, a self-consistent explanation of their radial profile and spherically-symmetric morphology remains elusive due to the difficulty in confining high-energy electrons and positrons within $\sim$20~pc regions of the interstellar medium. One proposed solution utilizes anisotropic diffusion to confine the electron population within a ``tube" that is auspiciously oriented along the line of sight. In this work, we show that while such models may explain a unique source such as Geminga, the phase space of such solutions is very small and they are unable to simultaneously explain the size and approximate radial symmetry of the TeV halo population.
\end{abstract}

\maketitle

\section{Introduction}

TeV halos are a new class of high energy $\gamma$-ray sources that are powered by pulsars~\cite{Linden:2017vvb}. The primary observational characteristics of TeV halos include: (1) a hard $\gamma$-ray spectrum consistent with inverse-Compton (IC) scattering, (2) a roughly spherically symmetric emission morphology that does not trace Galactic gas, (3) a coincidence with young and middle-aged pulsars, though millisecond pulsars may also produce TeV halos, (4) diffusive particle propagation that extends out to $\sim$10--50~pc~\cite{Abeysekara:2017hyn, Hooper:2018fih, Fang:2018qco, Tang:2018wyr, Profumo:2018fmz,  DiMauro:2019hwn, Sudoh:2019lav, DiMauro:2019yvh, LHAASO:2021crt, Hooper:2021kyp, Sudoh:2021avj}. This final observation is noteworthy, because TeV halos are larger than pulsar wind nebulae (PWNe) and remain bright for a longer period than supernova remnants (SNRs), but are more compact than expected for particle propagation in the standard interstellar medium (ISM).

Our understanding of TeV halos hinges on one key question: are TeV halos produced in peculiar regions of the ISM that have \emph{pre-existing} conditions ripe for halo formation? Or, conversely, are TeV halos produced throughout the bulk of the ISM and powered by the natal SNR, PWN, or potentially supergiant star which produces a local environment necessary for halo formation? In the former case, only a small fraction of pulsars will produce observable TeV halos. In the latter, TeV halos are expected to surround most energetic pulsars.

Observations support the latter case. Ref.~\cite{Linden:2017vvb} ranked all ATNF catalog~\cite{Manchester:2004bp} pulsars in terms of their ``spin-down flux" (spin-down power divided by the pulsar distance squared). Assuming that all pulsars convert an equivalent fraction of their spin-down power into $\gamma$-ray emission, the TeV halo flux should be proportional to spin-down flux. Indeed, Ref.~\cite{Linden:2017vvb} found that five of the seven pulsars with the highest spin-down-flux were detected as TeV sources, while none of the 48 dimmer objects in the HAWC field of view were detected. Subsequent observations detected TeV emission from another relatively high (the 11th brightest) pulsar~\cite{2017ATel10941....1R}. Additional TeV halo observations by the HAWC, H.E.S.S. and LHAASO collaborations have provided additional support for the conclusion that TeV halos can be found throughout the Milky Way~\cite{2018ATel12013....1B, Sudoh:2019lav, LHAASO:2021crt, Fang:2021qon}.

The theoretical arguments for ``innate" or ``source-produced" turbulence are less certain. Early studies focused on the potential that Geminga and Monogem exist in a ``low-diffusion pocket", that would potentially extend all the way to the solar position~\cite{HAWC:2017kbo}. However, such a model is incompatible with both local observations, which indicate that the diffusion coefficient near Earth is not abnormally low~\cite{Hooper:2017tkg}, and global observations, because the existence of many large low-diffusion regions surrounding pulsars would be incompatible with observed CR secondary-to-primary~\cite{Hooper:2017gtd, Fang:2018qco, Profumo:2018fmz}.

Several classes of models have been proposed to explain TeV halos. One popular model focuses on the potential for CRs accelerated by the pulsar or associated SNR~\cite{Fang:2019iym} to excite a resonant streaming instability that self-confines the CRs near the source~\cite{Evoli_Tim_2018, Mukhopadhyay:2021dyh}. These models potentially explain the evolution of halos, but many complexities of CR turbulence must be solved to make precise predictions. Rectilinear propagation models argue that diffusion is not inhibited, but particle propagation is instead ballistic on small scales, which produces an effective suppression of high-angle emission~\cite{Recchia:2021kty}. However, such models may require an unphysically high efficiency for the pulsar e$^+$e$^-$ production~\cite{Bao:2021hey}.

Finally, another kind of model argued that the apparent angular size of Geminga and Monogem are consequences of anisotropic diffusion with a maximal diffusion constant similar to the galactic average. In this scenario, the direction of efficient diffusion is oriented along the line-of-sight (LoS) towards Earth, while diffusion is strongly inhibited in the two visible directions perpendicular to the LoS~\cite{Huirong_anisotropic} (see also Ref.\cite{KaiYan_2022}). This model is theoretically motivated by synchrotron polarization measurements which indicate that local diffusion is dominated by flux tubes on scales between 1--100~pc~\cite{Haverkorn:2008tb, 2013A&A...558A..72I}. However, such a model does not predict that many TeV halos would be seen, as observable halos would only be expected from sources that have flux tubes that are fortuitously aligned towards Earth.

In this paper, we systematically re-examine the class of anisotropic diffusion models. We show that they cannot simultaneously account for the radial size and approximate spherical symmetry of the observed TeV halo population. We note that this conclusion holds for any CR-powered source (hadronic or leptonic), implying more generally that anisotropic diffusion does not dominate the propagation of particles near energetic sources.

\section{Anisotropic diffusion around PWN\MakeLowercase{e}}
\label{sec:models}

\subsection{Theory}\label{sec:models_th}
To study the lepton distribution, $u(\bm{r}, t, E_e)$, around pulsars, we make use of the standard transport equation~\cite{1964ocr..book.....G}:
\begin{equation}\label{eq:transport_eq_general}
    \frac{\partial u}{\partial t} = \frac{\partial}{\partial x_i} \left( D_{ij} \frac{\partial u}{\partial x_j} \right) - \frac{\partial}{\partial E_e} \left( \frac{\partial E_e}{\partial t} \, u \right) + \mathcal{S}(\bm{r},t, E_e)
\end{equation}
where $D_{ij}$ is the diffusion tensor, the energy derivative accounts for the energy losses and $\mathcal{S}(\bm{r}, t, E_e) = Q(E) L(t) \delta(\bm{r})$ is the source term, representing the flux injected by a point source located at $\bm{r}$~=~0 as a function of time. 

The diffusion of charged particles depends on the local magnetic field, $\bm{B}_{\mathrm{tot}}$. Magnetic fields in astrophysical plasmas can be described as $\bm{B}_{\mathrm{tot}} = \bm{B}_0 + \delta \bm{B}$, namely the sum of a large-scale background field $(\bm{B}_0)$, with a coherence scale between $\sim1-100\, \mathrm{pc}$~\cite{Strong:2007nh}, and a small-scale field $(\delta \bm{B})$ that depends on the size of the source that is powering turbulence. For TeV halos, typical turbulence scales are $L_{\mathrm{inj}} \sim \mathcal{O}(10) \, \mathrm{pc}$, while SNRs may inject turbulence up to $L_{\mathrm{inj}} \sim \mathcal{O}(100) \, \mathrm{pc}$.

To account for the effect of the magnetic field structure in particle transport, it is useful to decompose the diffusion tensor into directions parallel and perpendicular to the large-scale magnetic field lines as \mbox{$D_{ij} = D_\perp \delta_{ij} + \left(D_\parallel - D_\perp \right)b_i b_j$}, where \mbox{$b_{i} \equiv B_{i} \big/ |\bm{B}_0|$} for \mbox{$i,j = x,y,z$}, in a Cartesian reference frame. Placing the background magnetic field along the $\bm{z}$-axis \mbox{$\left(\bm{B}_0 = \left( 0, 0, B_0 \right)\right)$} we exploit the axisymmetric nature of the problem and write \mbox{$D_{xx} = D_{yy} = D_\perp$}, \mbox{$D_{zz} = D_\perp + \left( D_\parallel - D_\perp \right) \, B^2_z \big/ |B_0|^2 = D_\parallel$}, and all \mbox{$D_{ij} = 0$}. This allows us to solve Equation~\eqref{eq:transport_eq_general} in cylindrical coordinates $(r, z, \phi)$, for a cylinder oriented along the $\bm{z}$ axis, such that $D_{zz} = D_{\parallel}$ and $D_{rr} = D_\perp$, where $r$ is the polar coordinate $\sqrt{x^2 + y^2} = r$. The diffusion equation becomes:
\begin{equation}\label{eq:diffeq}
\begin{split}
\frac{\partial u}{\partial t} (r,z,t,E_e) = \frac{1}{r}\frac{\partial}{\partial r}\left( r D_\perp \frac{\partial u}{\partial r} (r,z,t,E_e) \right)\\ + \frac{\partial}{\partial z}\left( D_\parallel \frac{\partial u}{\partial z} (r, z, t, E_e) \right) + \\ + \frac{\partial }{\partial E_e} \left( \frac{\partial E_e}{\partial t} \, u (r, z, t, E_e) \right) + \mathcal{S}(r,z,t,E_e)
\end{split}
\end{equation}
where, because of the cylindrical symmetry, the gradients involving the azimuthal coordinate $\phi$ vanish. 

The physics behind parallel and perpendicular diffusion on scales similar to the Larmor radius is different~\cite{1966ApJ...146..480J}. Parallel diffusion is the result of the scattering of particles against $\delta \bm{B}$, while (mainly) the random walk of the lines themselves (\textit{field-line random walk}) is responsible for perpendicular diffusion. Therefore, if the injected turbulence is strong enough to considerably affect the preferential direction of the background field $\bm{B}_0$ on small scales, particle motion tends not to have a privileged direction, and is instead isotropic. Conversely, a weak turbulence does not alter the direction of $\bm{B}_0$. The intensity of the injected turbulence is represented by the so-called \textit{Alfvénic Mach number}, defined as $M_A \approx \left( \delta B \big/ B_0 \right)\big\vert_{L_{\mathrm{inj}}}$ at the turbulence injection length-scale, $L_{\mathrm{inj}}$. 

Anisotropic diffusion is an enticing explanation for the TeV halo morphology because it can explain the spatial extension of halos without invoking a diffusion coefficient that is orders of magnitude below standard ISM diffusion~\cite{Abeysekara:2017hyn}. Whenever the background magnetic field direction is oriented with our LoS, we observe the TeV halo in only the directions where diffusion is inhibited, and the low-diffusion coefficient becomes a projection effect~\cite{Huirong_anisotropic}. However, the overall diffusion coefficient, which is uninhibited along the LoS, remains consistent with global cosmic-ray measurements.

Quantitatively, we set $D_\parallel$ to match cosmic-ray measurements (\textit{e.g.} the boron-over-carbon ratio) and set the perpendicular diffusion coefficient using the model \mbox{$D_\perp = D_\parallel M_A^4$} -- derived by Ref.~\cite{Yan_Lazarian_2008} for particle energies below $\sim 10 \, \mathrm{TeV}$, corresponding to Larmor radii smaller than the injection scale $L_{\mathrm{inj}} \sim \mathcal{O}(10-100) \, \mathrm{pc}$. TeV halo observations constrain our models to $0 < M_A \leq 1$, which spans from the anisotropic case $(\delta B \ll B_0$, or $M_A \simeq 0.1)$ to the isotropic one $(\delta B = B_0$, or $M_A = 1)$. This implies that particle diffusion perpendicular to the local field can be strongly inhibited, depending on the turbulence strength and injection scale.

We parameterize the energy scaling of parallel diffusion as \mbox{$D_{\parallel} = D_0 \left( E \big/ E_0 \right)^{\delta}$}, where $D_0$ 
is set at a chosen normalization energy $E_0$ and $\delta$ is derived from the spectral index of the turbulent power spectrum. 
We fix $D_0 = 3.8 \times 10^{28} \, \mathrm{cm^2 \, s^{-1}}$ at $E_0 = 1 \, \mathrm{GeV}$ and consider a Kolmogorov spectrum for which $\delta = 0.33$. We note that these values are standard~\cite{Trotta2011, Luque:2021nxb} and compatible with the first-principle calculations in Ref.~\cite{Fornieri:2020wrr}.

Leptons in the halo interact with their environment to produce bright $\gamma$-ray emission, predominantly through inverse-Compton scattering (IC) of the surrounding Interstellar Radiation Field (ISRF)~\cite{HAWC:2017kbo}. The IC emissivity results from the convolution of the photon field and the CR spectrum~\cite{IC_sigma}:
\begin{equation}
\label{eq:IC_th}
    \begin{split}
    \epsilon_{\mathrm{IC}}(E_{\gamma}, \bm{r}) = 4\pi\int dE_{\mathrm{ph}} \, 
    \frac{dn_{\mathrm{ph}}}{dE_{\mathrm{ph}}}(E_{\mathrm{ph}}, \bm{r}) \\ \times \int dE_e \, \frac{d\sigma_{\mathrm{IC}}}{dE_{\gamma}}(E_e, E_{\mathrm{ph}}, E_{\gamma}) \, \Phi_e(E_e, \bm{r})
    \end{split}
\end{equation}
where $d\sigma_{\mathrm{IC}} \big/ dE_\gamma$ is the differential production cross-section of $\gamma$-rays with energy $E_\gamma$ resulting from the collision of an electron with energy $E_e$ and a background photon with energy $E_{\mathrm{ph}}$, $dn_{\mathrm{ph}} \big/ dE_{\mathrm{ph}}$ is the spectral density of ISRF photons and $\Phi_e$ is the differential electron flux, which, for isotropic emission, is $\Phi_e(E_e, \bm{r}) = c \big/ 4\pi \times u(E_e, \bm{r})$. The differential cross section for IC is given by~\cite{IC_sigma}:
\begin{equation}
\begin{aligned}
&\frac{d\sigma_{\mathrm{IC}}}{dE_{\gamma}}(E_e, E_{\mathrm{ph}}, E_{\gamma}) = \frac{3\sigma_{\mathrm{T}} m_e^2}{4 E_{\mathrm{ph}} E_e^2} \\
&\times \, \left[2q\,\log(q) + (1+2q)(1-q) + \frac{(pq)^2(1-q)}{2(1+pq)}\right]
\end{aligned}
\end{equation} 
where $\sigma_{\mathrm{T}}$ is the Thomson cross section, $m_e$ is the electron mass. Here $p = 4E_{\mathrm{ph}}E_e \big/ m_e^2$ and
\begin{equation*}
q = \frac{E_{\gamma}m_e^2}{4E_{\mathrm{ph}}E_e^2 \left(1 - E_{\gamma} \big/ E_e \right)},
\end{equation*} 
where the cross-section vanishes outside $m_e^2 \big/ 4E_e^2 \leq 1$.

\begin{figure*}[t]
\hspace{-0.3cm}
    \includegraphics[width=1.00\linewidth]{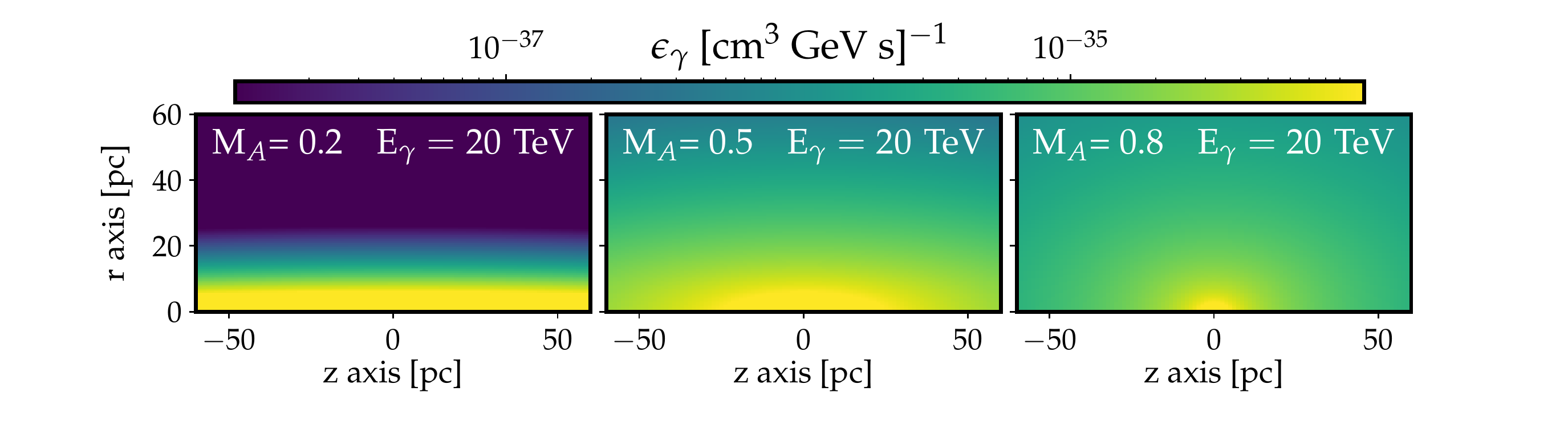}
    \includegraphics[width=0.92\linewidth]{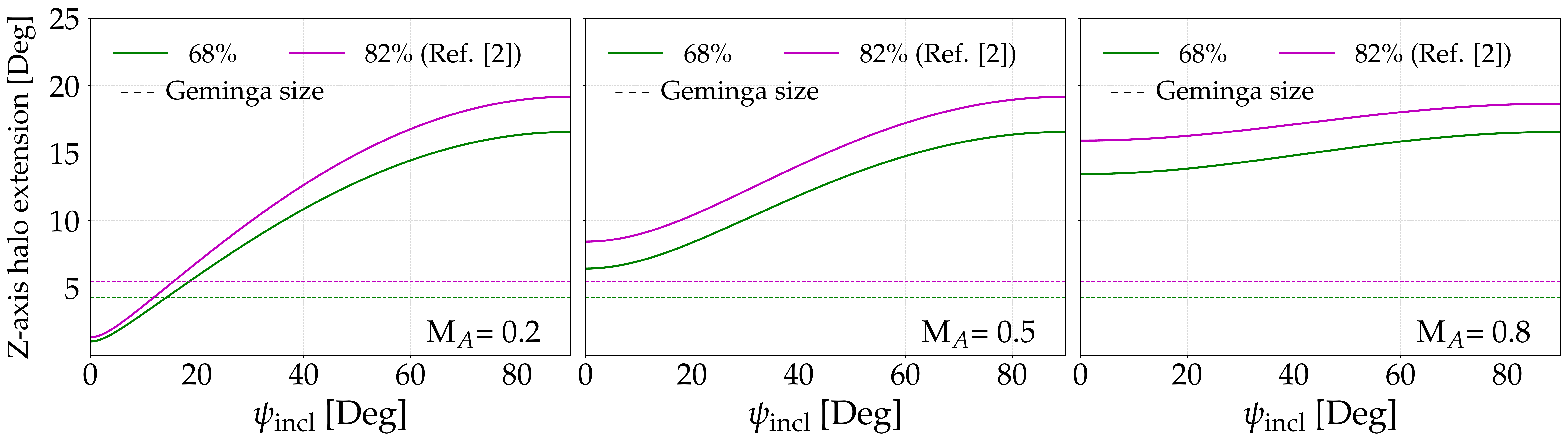}
    
    \caption{\textbf{Top panels:} $\gamma$-ray emissivity maps for different levels of anisotropic emission ($M_A = 0.2$, $M_A = 0.5$ and $M_A = 0.8$) at \mbox{$E_{\gamma} = 20$~TeV}. \textbf{Bottom panels:} TeV halo extension projected on the plane-of-sky, along the $Z$-axis for different inclination angles ($\psi_{\mathrm{incl}}$), compared to the Geminga's TeV halo size for $\sim68\%$ and $\sim82\%$ of the flux contained around the source (green and magenta dashed lines, respectively). A simulation with a larger window ($120$~pc) has been used to correctly compute the total extension of the halo.}
\label{fig:maps}
\end{figure*}

\subsection{Numerical setup}
\label{sec:models_su}
\vspace{-0.2cm}
Equation~\eqref{eq:diffeq} cannot be solved analytically. In this paper, we numerically evolve our system using a Crank-Nicolson scheme~\cite{crank_nicolson_1947} as detailed in Appendix~\ref{sec:AppA}. We examine values of $M_A$ spanning from $0.1$ to $1$ and align the large-scale magnetic field with the $\bm{z}$-axis. The diffusion equation is evolved on a 2D grid of radius $60$~pc and height $[-60 \, \mathrm{pc}, \, +60 \, \mathrm{pc}]$ with $(N_r, \, N_z) = (200, \, 200)$ points. Adopting a reference distance to Geminga of $d_{\mathrm{Gem}} = 250 \, \mathrm{pc}$, this corresponds to a window $\Delta \phi \simeq [-13^\circ, \, +13^\circ]$ and angular resolution $\phi_{\mathrm{res}} \simeq 0.1^\circ$.

We assume that the leptons are injected from a point-like pulsar source, noting that the assumed source size does not affect our results. Drawing on pulsar observations, we set the luminosity function to $L(t) = L_0 \times (1 + t/\tau_0)^{-\frac{n+1}{n-1}}$, where $L_0$ is the luminosity of the source at $t=0$, $n$ is the braking index and $\tau_0$ is the pulsar spin-down timescale. In our simulations we set $L_0= 2.8\times10^{37}\, \mathrm{erg \, s^{-1}}$, $\tau_0=12 \, \mathrm{kyr}$ and $n=3$, and normalize our results by imposing that the total energy released by the pulsar since its birth is $W_e=1.1\times10^{49}\, \mathrm{erg}$, consistent with previous studies~\cite{recchia2021does, Huirong_anisotropic, DiMauro_Leptons, Evoli_Tim_2018}.

We convert this spin-down power into electron and positron pairs with an efficiency $\eta$, which is fit to data, but cannot exceed 1, yielding an injection spectrum \mbox{$Q(E_e) = \eta \, Q_0 \left( E_e \big/ \mathrm{GeV} \right)^{-\alpha} \times e^{-E_e/E_{\mathrm{cut}}}$}, where \mbox{$\alpha = 1.6$~\cite{Hooper:2017gtd}} and $E_{\mathrm{cut}} = 200 \, \mathrm{TeV}$, and $Q_0$ is a normalization constant. We compute the electron flux from $0.1 - 300 \, \mathrm{TeV}$ and the IC-produced $\gamma$-ray flux from $0.1 - 200 \, \mathrm{TeV}$. In our setup, we stop the simulation at the age of Geminga, $t_{\mathrm{ch}} \sim 342 \, \mathrm{kyr}$. 

We point out that, by comparing the different relevant timescales, we can conclude that our final result is robust with respect to the age of the pulsar. There are three relevant timescales: (1) the age of the pulsar, (2) the spin-down timescale of the pulsar, which we compute by determining the time period over which the pulsar luminosity changes by a factor of $e$, and (3), the diffusion timescale, which determines the rate at which the leptons produced by the pulsar leave the simulation volume. The age of the pulsar in our simulation is 342~kyr. We note that the pulsar spindown timescale is a power-law, and not an expontential, so it changes significantly as a function of pulsar age. At 342~kyr, the effective pulsar spindown timescale is $\sim$180~kyr. The diffusion timescale, on the other hand, is only $\sim$2~kyr (for $D_0=3.8\times10^{28}$cm$^2$/s). Over this timescale, the spin-down power only changes by $\sim 1\%$. This means that, while the spindown timescale affects the normalization of the particle density, it will have no effect on the morphology of the TeV halo in our simulation. 


In fact, the axial ratio of diffusion parallel and perpendicular to the magnetic field lines is even more robust to changes in the pulsar age or spin-down timescales, because it is based on the ratio for particles to diffuse in each direction, which is independent of the instantaneous pulsar power. Thus, changes in the age of the modeled pulsar will produce negligible changes in the results of our study.
On top of this, we also stress that the injection energy dependence assumed does not affect any of our conclusions on the morphology and radial profile of the $\gamma$-ray emission.

Synchrotron and IC energy losses are calculated as:
\begin{equation}
    \frac{\partial E_e}{\partial t} = -\frac{4}{3}\, c \, \sigma_{\mathrm{T}} \left[ U_B + f^i_{\mathrm{KN}}(E_e) \, U_i \right] \left(\frac{E_e}{m_e c^2} \right)^2,
\end{equation}
where $\sigma_{\mathrm{T}} \simeq 6.65 \times 10^{-25} \, \mathrm{cm}^2$ is the Thomson cross-section, and $(U_B, \, U_i)$ are the magnetic field and ISRF energy densities. We set $U_B$ to be equal to the dominant ordered field, $U_B = B^2_0 \big/ 4 \pi \simeq 0.22 \, \mathrm{eV \, cm^{-3}}$ for $B_0 = 3 \, \mu \mathrm{G}$, and calculate the ISRF using a six-component blackbody model consisting of CMB, infrared, optical and UV components with temperatures \mbox{$T_i = [2.6, 33.1, 313, 3250, 6150, 23200]$~K} and energy densities \mbox{$U_i = [0.26, 0.25, 0.055, 0.37, 0.23, 0.12]$~eV~cm$^{-3}$}~\cite{Carmelo_KN}. For each component we calculate the Klein-Nishina suppression factor, $f^i_{\mathrm{KN}}$, following Equation~\eqref{eq:KN_fact}~\cite{2010NJPh...12c3044S}, an approximation that is valid for the energy-range and accuracy we require, but may fail for high-precision GeV measurements~\cite{DiMauro:2020cbn}.

\section{Results}
\label{sec:results}

Using the modelling described above, we produce mock observations for Geminga-like TeV halos for various quantities of the parameters $M_A$ and the inclination angle of the simulation with respect to the LoS, $\psi_{\mathrm{incl}}$. We note that these models are produced in a 2D simulation with dynamics that are dependent on the parameter $M_A$. The parameter $\psi_{\mathrm{incl}}$, on the other hand, corresponds to mock observations taken from different angles with respect to the axes of the simulation. We utilize the lower-case letters $r$ and $z$ when we discuss the physical coordinates of the simulation, and the upper-case letters $R$ and $Z$ to denote the projected coordinate system along our line of sight (see Figure~\ref{fig:propagated_cylinder}). We note that $R$~=~$Z$ when $\psi_{\mathrm{incl}}$~=~0$^\circ$ and that $R$~=~$r$, $Z$~=~$z$ when $\psi_{\mathrm{incl}}$~=~90$^\circ$.

Figure~\ref{fig:maps} (top) shows the morphology of the $\gamma$-ray emissivity as a function of $r$ and $z$ at $20$~TeV, as computed in Equation~\ref{eq:IC_th}. In Figure~\ref{fig:maps} (bottom), we show the \emph{observed} extension of the simulated halo as a function of the angle $\psi_{\mathrm{incl}}$, which corresponds to rotations of our simulated cylinder with respect to our LoS along the $r$-axis, and compare our results to the 68\% and 82\% of the flux contained in the Geminga TeV halo as reported by Ref.~\cite{Abeysekara:2017hyn}. We note that rotations around the $\bm{z}$-axis do not change the morphology of the halo with respect to our LoS due to the cylindrical symmetry of the system, while rotations around the $\bm{r}$-axis change the morphology that is projected on the plane-of-the-sky ({\it c.f.} Figure~$2$ of Ref.~\cite{Huirong_anisotropic}).

Figure~\ref{fig:maps} demonstrates that if anisotropic diffusion produces TeV halos, we should detect a variety of both highly extended and asymmetric objects (as seen at different inclination angles, $\psi_{\mathrm{incl}}$). This is in tension with the fact that observed TeV halos have similar sizes and approximate spherical symmetry. The model is constrained from two directions: (i) for values of $M_A \leq 0.5$, the asymmetry of each TeV halo becomes pronounced and observations would show ``ovals" or "strings" in the TeV sky, while spherically symmetric halos would be observed only when $\psi_{\mathrm{incl}} \sim 0^{\circ}$. (ii) for values of $M_A \geq 0.5$ the halo appears roughly spherically symmetric, but the lack of inhibited diffusion makes the halo too large to explain observed systems. Notably, we see that for $M_A \geq 0.5$ there is no value of $\psi_{\mathrm{incl}}$ for which the containment angle along the $\bm{z}$-axis is consistent with HAWC observations of Geminga.

We can formalize the excluded $\psi_{\mathrm{incl}}$ angles based on the morphology and symmetry of simulated TeV halos by imposing two conditions: (i) that the emission should not be very asymmetric (\textit{i.e.} the extension of the halo in any direction should not be much larger than the extension in the perpendicular one). (ii) the size of the emission should not be much larger than $5.5^{\circ}$ (\textit{i.e.} $24$~pc, given the distance from Geminga), to be consistent with the size of Geminga reported in Ref.~\cite{Abeysekara:2017hyn}, which corresponds to $\sim82\%$ of the flux contained around the source. We additionally calculate the size the halo at $\sim68\%$ ($\sim1\sigma$) containment, which is $4.3^{\circ}$ ($\sim19$~pc) for Geminga.

To quantify the first condition (hereafter, the \textit{symmetry condition}) we impose that the projected extension of the halo in one direction must not be more than $100\%$ larger than in the other direction ($Z/R < 2$), which is a very conservative choice. 
Figure~\ref{fig:s_condition} of the supplementary material (SM) shows the $Z/R$ ratio for different inclination angles. The second condition (hereafter, the \textit{size condition}, see bottom panels of Figure~\ref{fig:maps}) imposes that the extension of the halo projected on the plane-of-sky along $Z$ is within the size uncertainty reported by HAWC, which is $5.5\pm0.7^{\circ}$ ($\sim24\pm3$~pc). While the first condition only depends on the ratio $D_{\bot} \big/ D_{\parallel} = M_A^4$, the second depends on both such ratio and the normalization of $D_{\parallel}$, which is fixed to the diffusion coefficient obtained from analyses of CR secondary-to-primary ratios. Since the normalization of the $D_{\parallel}$ in the Galaxy is uncertain by at least $\sim 30\%$, mainly due to cross sections uncertainties~\cite{DeLaTorreLuque:2021yfq, Luque:2021ddh, Korsmeier:2021brc}, we have also tested other values of the normalization of $D_{\parallel}$ around $D_0 = 3.8 \times 10^{28} \, \mathrm{cm^2 \, s^{-1}}$, as we discuss below.

\begin{figure}[!t]
\centering
\includegraphics[width=\linewidth]{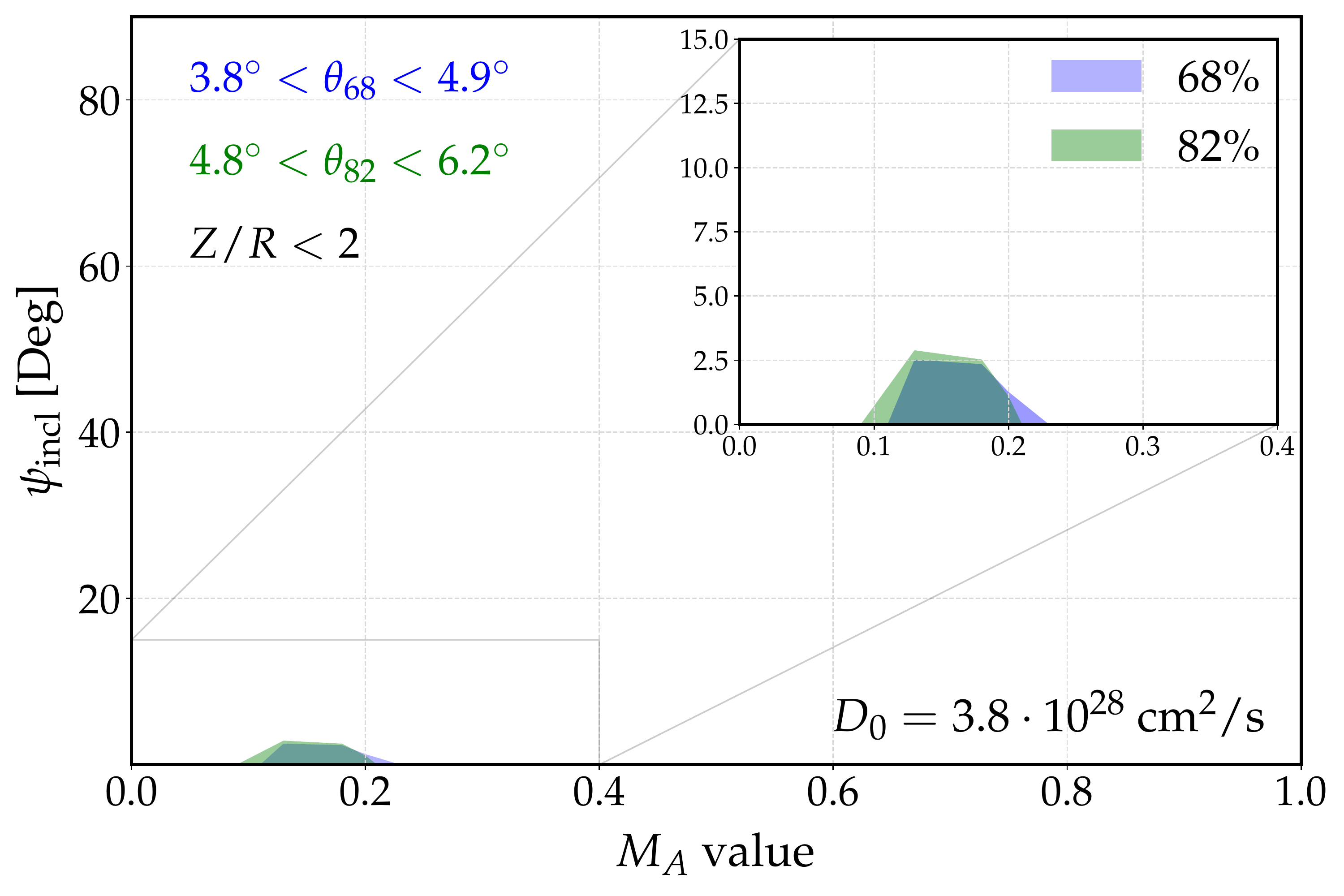}
\caption{The parameter space of inclination angles ($\psi_{\mathrm{incl}}$) and asymmetric diffusion parameters which produce TeV halos that fulfill the symmetry ($Z/R < 2$) and size ($\theta_c \sim  \theta_\mathrm{Geminga} \pm 1\sigma$) conditions at both 68\% and 82\% containment.}
\label{fig:Allowed_space}
\end{figure}

In Figure~\ref{fig:Allowed_space}, we show the constraint on the TeV halo population in the parameter space of $M_A$ and $\psi_{\mathrm{incl}}$. Only a very reduced space of inclination angles ($\psi_{\mathrm{incl}} < 5^{\circ}$) is able to simultaneously account for the radial size and measured symmetry of a typical TeV halo. 
This means that, unless there is a reason to believe that all existing TeV halos are aligned with our LoS, the anisotropic model is not able to explain the observation of multiple symmetric TeV halos and the lack of observed asymmetric ones. Quantitatively, we note that if all inclination angles are equally probable, then the probability of observing a TeV halo with an inclination angle less than 5$^\circ$ is given by $P$~=~[cos(0$^\circ$) - cos(5$^\circ$)] \big/ [(cos(0$^\circ$) - cos(90$^\circ$)], and the probability of finding 5 of the 11 brightest TeV halo candidates in this region would be given by:

\begin{equation*}
{11 \choose 5} \, P^5 \, (1-P)^6 =  3.6 \times 10^{-10}.
\end{equation*}

In Figure~\ref{fig:D0+-} of the appendix, we show the resulting TeV halo parameter space for values of $D_{\parallel}$, $D_0$, covering a one order of magnitude (from $10^{28}$ to $10^{29} \, \mathrm{cm^2 \, s^{-1}}$). As expected, larger values of $D_0$ further reduce the allowed parameter space, because the halo becomes more extended, while lower $D_0$ values only slightly increase the fraction of inclination angles that satisfy both conditions, even for our extreme values that are in significant tension with galactic secondary-to-primary ratios if these values are standard for the Galaxy.

\begin{figure*}[!th]
\includegraphics[width=1.\linewidth]{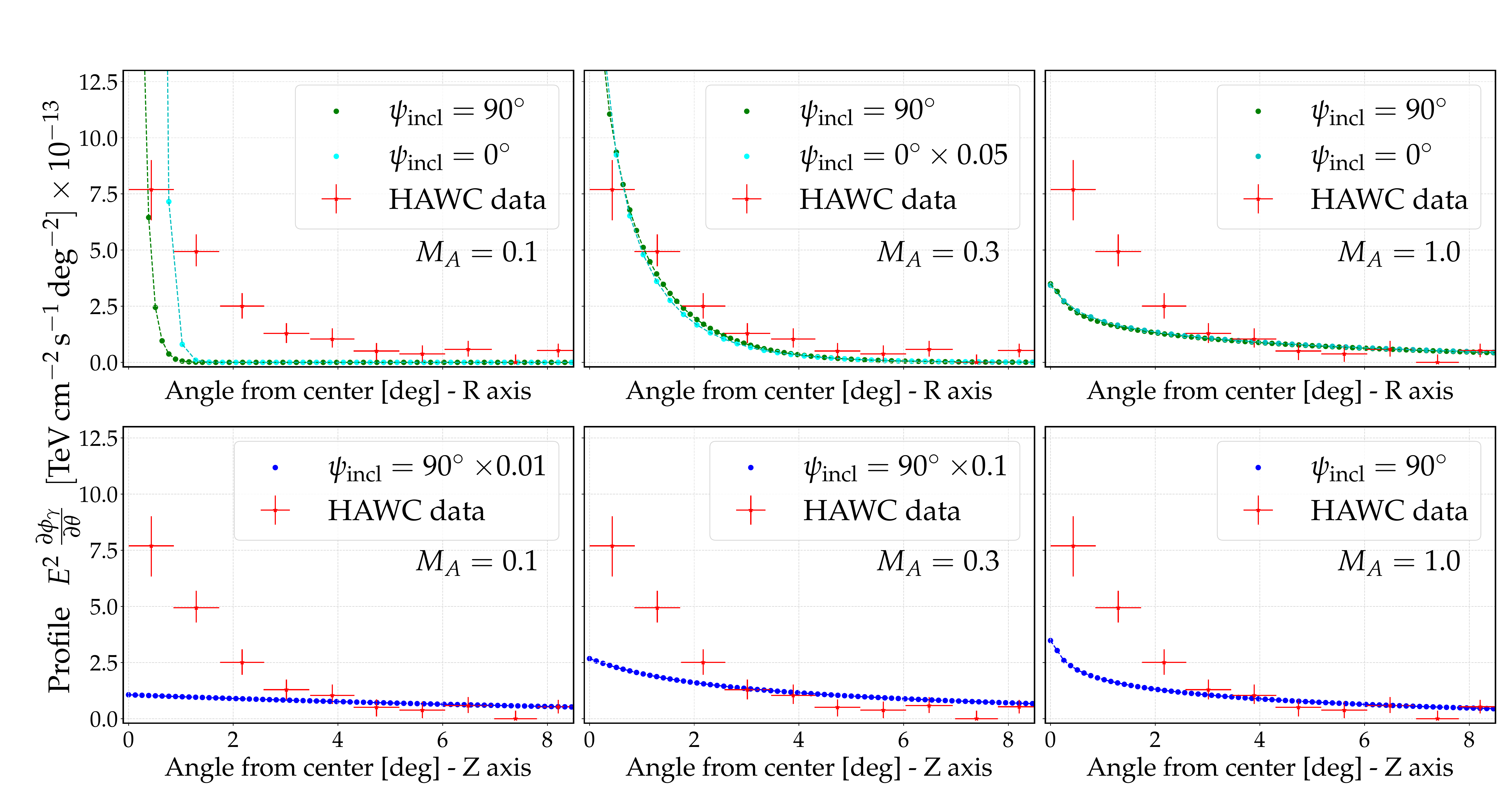}
\caption{Gamma-ray surface brightness for different values of $M_A$ ($M_A = 0.1$, $0.3$, $1$) at $\psi_{\mathrm{incl}}=90^{\circ}$ and  $\psi_{\mathrm{incl}}=0^\circ$, compared to the HAWC surface brightness. \textbf{Top panels}: gamma-ray emission integrated along $R$-axis. \textbf{Bottom panels}: gamma-ray emission along $Z$-axis. The emission is integrated for gamma-ray energies from $5$ to $50$~TeV. For each $M_A$, the intensity is scaled by the number shown in the legend. There is no $\psi_{\mathrm{incl}}$~=~0$^\circ$ line in the $Z$-case: such axis would be aligned with our LoS and thus the extension is not observed.}
\label{fig:los}
\end{figure*}

The relatively simple \emph{symmetry} and \emph{size} conditions already rule out the vast majority of the $M_A \big/ \psi_{\mathrm{incl}}$ parameter space. 
Additionally, the integrated profile is expected to show clear signatures of anisotropic diffusion when performed following independently perpendicular axes (i.e. the integrated $\gamma$-ray emission along the different axes is expected to be different). These signatures would be detectable, although no collaboration reported this kind of differences yet. In fact, both the results from Fig.~\ref{fig:Allowed_space} and the integrated profile are complementary information that must be used to prove or discard the observation of asymmetric TeV halos.

In Figure~\ref{fig:los}, we show the emission profiles in both the $\bm{r}$ (perpendicular to the magnetic field) and $\bm{z}$ (parallel to the field) directions for values of $M_A = 0.1$, $M_A = 0.3$ and $M_A = 1$. These are computed integrating the emissivity obtained from Eq.~\ref{eq:IC_th} along our LoS at each projected axis. Given the anisotropic structure of the predicted halos, the profile is computed along the $Z$ and $R$ axes separately. 
A crucial point here is that the computation of the profile by the HAWC collaboration was done taking the average emission at circular rings around the source, which does not allow the observation of any feature of anisotropy or asymmetry from the halo. Therefore, this kind of profile is not appropriate for asymmetric (anisotropic) objects. We point out that the recent works studying TeV halos compute their profile assuming circular symmetry as well~\citep{Huirong_anisotropic, Lopez-Coto:2017pbk, DiMauro:2019yvh, Fang:2021qon, recchia2021does}. However, we go beyond previous works and calculate the independent profiles in the $r$ and $z$ direction, which allows us to gauge the asymmetry of our model.
To have a qualitative comparison, HAWC's surface brightness~\citep{HAWC:2017kbo} is also shown in Fig.~\ref{fig:los}. As discussed, in the case of an asymmetric halo in the $\psi_{\mathrm{incl}}=90^{\circ}$ case, observations would detect a profile that is starkly different in each direction (at least for $M_A \leq 0.5$). This remains valid for angles $\psi_{\mathrm{incl}}>0^{\circ}$.


We stress that our results are not in contradiction with the results found by the authors of Ref.~\citep{Huirong_anisotropic}, since, in fact, the values of M$_A$ and $\psi_{Incl}$ that are compatible with our conditions are M$_A \lesssim 0.3$ and $\psi_\mathrm{incl} \lesssim 5^{\circ}$ Instead, our results indicate that the phase space for this solution is small, and the probability of having multiple systems in such a configuration is extremely low.

\section{discussion and conclusions}
\label{sec:conclusion}
\vspace{-0.3cm}
TeV halos constitute a new class of astrophysical objects which have the capability to significantly advance our understanding of galactic diffusion~\citep{Review}. In this work, we have demonstrated that one of the more popular models, where anisotropies in local diffusion explain the TeV halo morphology, is inconsistent with TeV halo observations. Specifically, we have explored and analyzed different morphological signatures of anisotropic diffusion that are predicted by this model but are not observed in detected TeV halos. 

We have analyzed the morphology of anisotropic TeV halos as a function of two key parameters: $M_A$, which controls the ratio of the diffusion coefficients perpendicular to and along the background magnetic field, and $\psi_{\mathrm{incl}}$, which controls the angle between the magnetic field and the observer's LoS. Our results constrain anisotropic TeV halo models in three ways: (1) we constrain $M_A$ to be smaller than $\sim 0.5$ to prevent the TeV Halos from becoming too large compared to current measurements, (2) we constrain $\psi_{\mathrm{incl}}$ to be less than $\sim 5^\circ$ in order to prevent observed TeV halos from having a significant visual asymmetry that would appear oblong or ``spaghetti shaped" on the sky, (3) we show that the expected surface brightness along different axes is significantly different for asymmetric objects, which could lead to easily discard values of $M_A$ smaller than $\sim 0.3$. 
In this context, we stress that it would be crucial to experimentally measure the integrated emission projected along the different axes, since such a test would be able to unequivocally detect signatures of anisotropic diffusion.

To be precise, our analysis specifically indicates that anisotropic diffusion cannot explain the observation of several TeV halos in any scenario where the diffusion coefficient in the uninhibited direction is compatible with best-fit values from galactic secondary-to-primary ratios. Our models leave open the possibility that the diffusion coefficient surrounding TeV halos is mildly anisotropic. However the diffusion coefficient in every direction must be significantly inhibited compared to the average diffusion coefficient of the Milky Way. As a result, TeV halos must occupy or generate regions with unique diffusion characteristics compared to the Milky Way average. We note that these conclusions are generically true for any CR source in the galaxy, and indicate that particle diffusion near sources with observable $\gamma$-ray emission cannot simultaneously be strongly anisotropic and have uninhibited diffusion along the preferential direction. Although our analysis is based on the anisotropic-diffusion model put forward in \citet{Yan_Lazarian_2008}, our conclusions remain valid for any anisotropic model where the scalings of the perpendicular and parallel diffusion coefficients are similar --- namely $\delta_{\parallel} \simeq \delta_{\perp}$, given the typical parameterization \mbox{$D_{\parallel, \perp} \propto E^{\delta_{\parallel}, \delta_{\perp}}$} --- which is supported by numerical simulations~\citep{Giacinti:2017dgt}.

In conclusion, observations of suppressed and spherically symmetric diffusion provide further credence in favor of models where the diffusivity is reduced not due to a geometrical effect, but rather \textit{intrinsically} inhibited/subdominant due to subtle mechanisms, either generated by the compact object or pre-existing in the region. These models include, for instance, (i) models with self-generated turbulence that efficiently confine CRs more than typical for the ISM~\citep{Evoli_Tim_2018, Fang:2019iym, Mukhopadhyay:2021dyh}, (ii) models with rectilinear propagation in the first stage of the particle injection~\citep{recchia2021does}, or (iii) models where the correlation length for the magnetic field is extremely small ($\sim 1 \, \mathrm{pc}$), such that particles are trapped within the magnetic field structure of the halo on timescales equivalent to HAWC-observations~\citep{Lopez_Coto_2018}.

\vspace{-0.4cm}
\section*{Acknowledgements}
\vspace{-0.4cm}
\noindent We thank Elena Amato, Mattia Di Mauro, Fiorenza Donato, Carmelo Evoli, Gwenael Giacinti, Ruoyu Liu, Ruben L{\'o}pez-Coto, and Alexandre Marcowith for helpful comments. PDTL and TL are supported by the Swedish National Space Agency under contract 117/19. TL is also partially supported by the Swedish Research Council under contract 2019-05135 and the European Research Council under grant 742104. This project used computing resources from the Swedish National Infrastructure for Computing (SNIC) under project Nos. 2021/3-42, 2021/6-326 and 2021-1-24 partially funded by the Swedish Research Council through grant no. 2018-05973.

\bibliography{main}

\onecolumngrid
\appendix
\section{Orientation of the halo structure with respect to our line-of-sight}
In this appendix, we report a sketch of the structure of the propagated particles, with the axes in capital letter $(\hat{Z}, \hat{R})$, referring to the reference frame of the central object. In particular, the figure represents the case of anisotropic diffusion ($M_A < 1$), when a clear cylindrical symmetry appears.
\begin{figure}[h!]\label{fig:propagated_cylinder}
    \centering
        \includegraphics[width=0.4\textwidth]{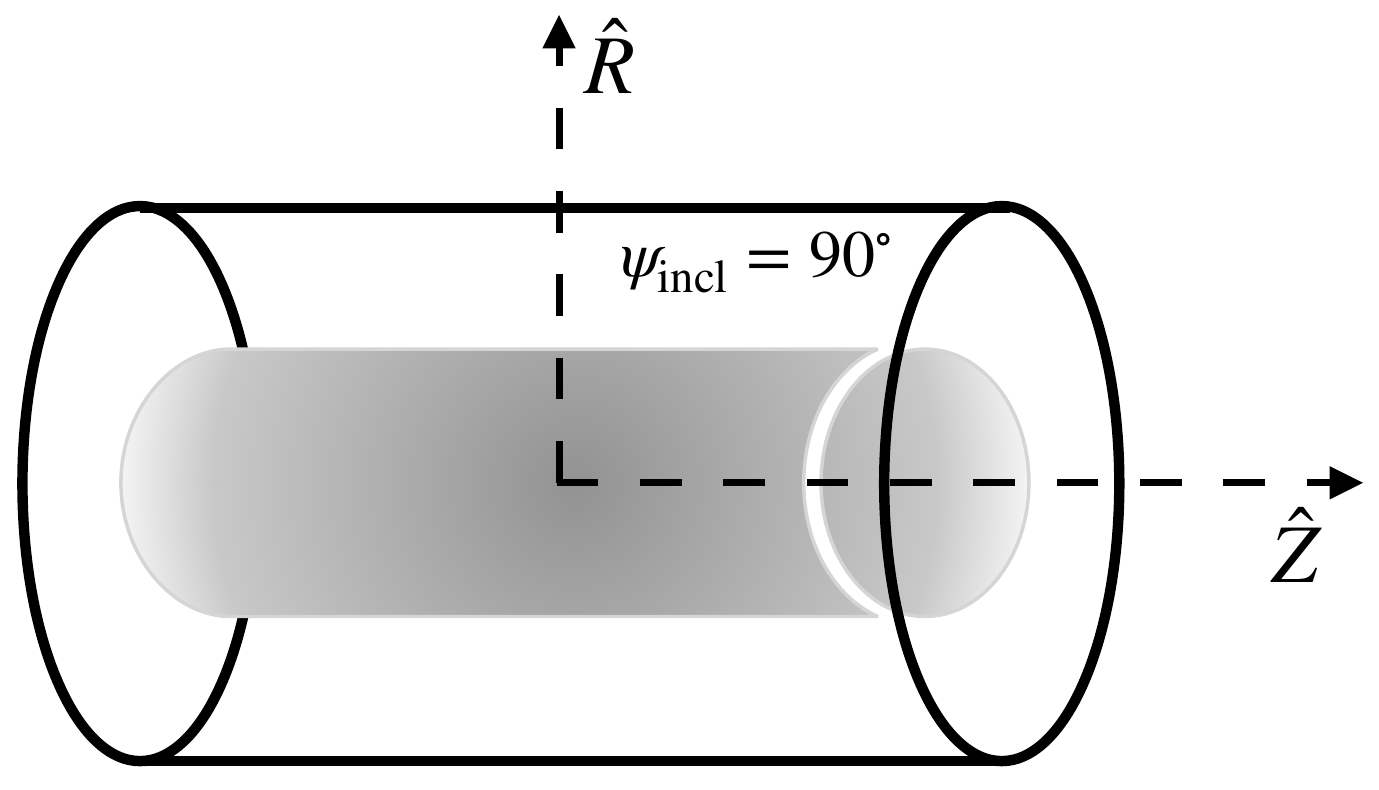}
        \hspace{0.9cm}
        \includegraphics[width=0.17\textwidth]{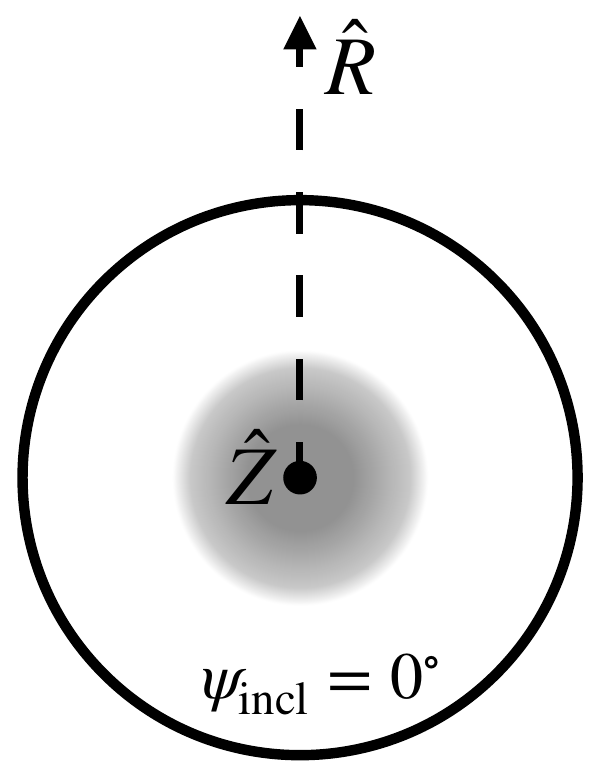}
    \caption{\small{Sketch of the cylindrical structure of the propagated particles, for the two reference inclinations used in the present work, with respect to our line of sight, $\psi_{\mathrm{incl}} = 90^\circ$ (left panel) and $\psi_{\mathrm{incl}} = 0^\circ$ (right panel). In the right panel, the $\hat{Z}$ is pointing out of the screen.}}
\end{figure}

\newpage
\section{Numerical solution of propagation equation}
\label{sec:AppA}
This appendix aims at detailing the numerical scheme implemented to solve the transport equation described in the text. We use the Crank-Nicolson (CN) expansion, which is second order in energy, space and time. A detailed description of our numerical discretization, an example script with this numerical prescription and a document clarifying some of the relevant numbers for the Geminga TeV halo based on Ref.~\citep{Abeysekara:2017hyn} are publicly available at \url{https://github.com/tospines/Analyses-and-plotting-codes/tree/main/Anisotropic_TeV_Halos}. The numerical algorithm for our scheme is given by:

\begin{equation}
\label{eq:eq_numeric}
\begin{split}
\frac{u^{\tau+1}_{i,k,\epsilon} - u^{\tau}_{i,k,\epsilon}}{\Delta t} = \frac{D_{\bot}}{\Delta r^2} \times \frac{1}{2} \left[ (u^{\tau}_{i+1 ,k,\epsilon} - 2u^{\tau}_{i,k,\epsilon} + u^{\tau}_{i-1,k,\epsilon} )  + (u^{\tau+1}_{i+1 ,k,\epsilon} - 2u^{\tau+1}_{i,k,\epsilon} + u^{\tau+1}_{i-1,k,\epsilon} ) \right] + \\ + \frac{D_{\bot}}{2r_{i}} \frac{(u^{\tau}_{i+1,k,\epsilon} - u^{\tau}_{i-1,k,\epsilon}) + (u^{\tau+1}_{i+1,k,\epsilon} - u^{\tau+1}_{i-1,k,\epsilon} )}{2\Delta r} + \\ + \frac{D_{\parallel}}{\Delta z^2} \times \frac{1}{2} \left[ (u^{\tau}_{i ,k+1,\epsilon} - 2u^{\tau}_{i,k,\epsilon} + u^{\tau}_{i,k-1,\epsilon} )  + (u^{\tau+1}_{i ,k+1,\epsilon} - 2u^{\tau+1}_{i,k,\epsilon} + u^{\tau+1}_{i,k-1,\epsilon} ) \right] + \\ + \frac{b(E_{\epsilon})}{2}\times\frac{(u^{\tau}_{i,k,\epsilon+1} - u^{\tau}_{i,k,\epsilon}) + (u^{\tau+1}_{i,k,\epsilon+1} - u^{\tau+1}_{i,k,\epsilon})}{\Delta E}
- \frac{1}{2} \left(u^{\tau}_{i,k,\epsilon} +  u^{\tau+1}_{i,k,\epsilon}\right) \frac{4}{3}\, \frac{c \, \sigma_{\mathrm{T}} \, U_B }{(m_ec^2)^2} \, 2 E_{\epsilon} + \\ - \frac{1}{2} \left(u^{\tau}_{i,k,\epsilon} + u^{\tau+1}_{i,k,\epsilon}\right) \frac{4}{3}\, \frac{c \, \sigma_{\mathrm{T}} U_{\mathrm{ph}}}{(m_ec^2)^2} \, 2 E_{\epsilon} \mathcal{F}_{\mathrm{KN}}(E_{\epsilon}) + \\  - \frac{1}{2} \left(u^{\tau}_{i,k,\epsilon} + u^{\tau+1}_{i,k,\epsilon}\right) \frac{4}{3}\, \frac{c \, \sigma_{\mathrm{T}} U_{\mathrm{ph}}}{(m_ec^2)^2} \left( E_{\epsilon}^2 \frac{\mathcal{F}_{\mathrm{KN}}(E_{\epsilon+1}) - \mathcal{F}_{\mathrm{KN}}(E_{\epsilon})}{\Delta E} \right)   + \\ + S(r_i, z_k, E_{\epsilon}, t^{\tau+1/2}) \, \, .
\end{split}
\end{equation}

\noindent Here $u$ is the density of electrons, $E$ represents their energy and the subscripts $i$, $k$ and $\epsilon$ are the indexes of the spatial step in $r$ and $z$ direction and energy bin, respectively, and $\tau$ is the time-step index. The diffusion coefficients in the perpendicular and parallel direction to the magnetic field (defined as the $\bm{z}$-axis) are denoted as $D_{\parallel}$ and $D_{\bot}$, respectively. $\sigma_{\mathrm{T}}=6.652\times 10^{-29}$~m$^2$ is the Thomson cross section, the energy density of the magnetic field is denoted as $U_B=\frac{B_0}{4\pi}$, the energy density of the different photon fields included in this study is $U_{\mathrm{ph}}= U_{\mathrm{CMB}} + U_{\mathrm{IR}} + U_{\mathrm{opt}} + U_{\mathrm{UV}}$ and the term $b=\frac{dE}{dt} = -\frac{4}{3} \frac{c \, \sigma_{\mathrm{T}} \left( U_B + U_{\mathrm{ph}} \cdot \mathcal{F}_{\mathrm{KN}}\right)E^2}{(m_ec^2)^2}$ is the term of synchrotron and inverse-Compton energy losses, where $c$ is the speed of light and $m_e$ is the electron mass. 
The Klein-Nishina factor $\mathcal{F}_{\mathrm{KN}} = (f^i_{\mathrm{KN}}, f^j_{\mathrm{KN}}, ...)$ is defined as in Ref.~\cite{2010NJPh...12c3044S}:
\begin{equation}
f^i_{\mathrm{KN}}(E_e) = \frac{\sigma_{\mathrm{KN}}}{\sigma_{\mathrm{T}}} \simeq \frac{\frac{45}{64 \pi^2} \times (m_e c^2/ k_{\rm B}T_i)^2}{ \frac{45}{64 \pi^2} \times  (m_e c^2/ k_{\rm B} T_i)^2 + (E^2 / m_e^2 c^4)},
\label{eq:KN_fact}
\end{equation}
where $T_i$ is the temperature of each ISRF component. Recently, an analysis by Ref.~\cite{DiMauro:2020cbn} showed that this approximation of the true Klein-Nishina cross-section is now insufficient to describe highly-precise AMS-02 data at the percent level. However, the accuracy of this fit is more than sufficient to describe the electron cooling parameters which are shown here.

To speed up the computations, we differentiate at first order over energy ($\mathcal{O}(dE)$), which provides sufficient accuracy so long as the spacing of our energy steps ($\delta E$) is larger than the energy-loss in a given time-step ($\delta t$). Our results are accurate to second order in time and space ($\mathcal{O}(dx^2\, \, dt^2)$).
\begin{figure}[!t]
    \centering
    \includegraphics[width=0.49\linewidth, height = 0.31\textwidth]{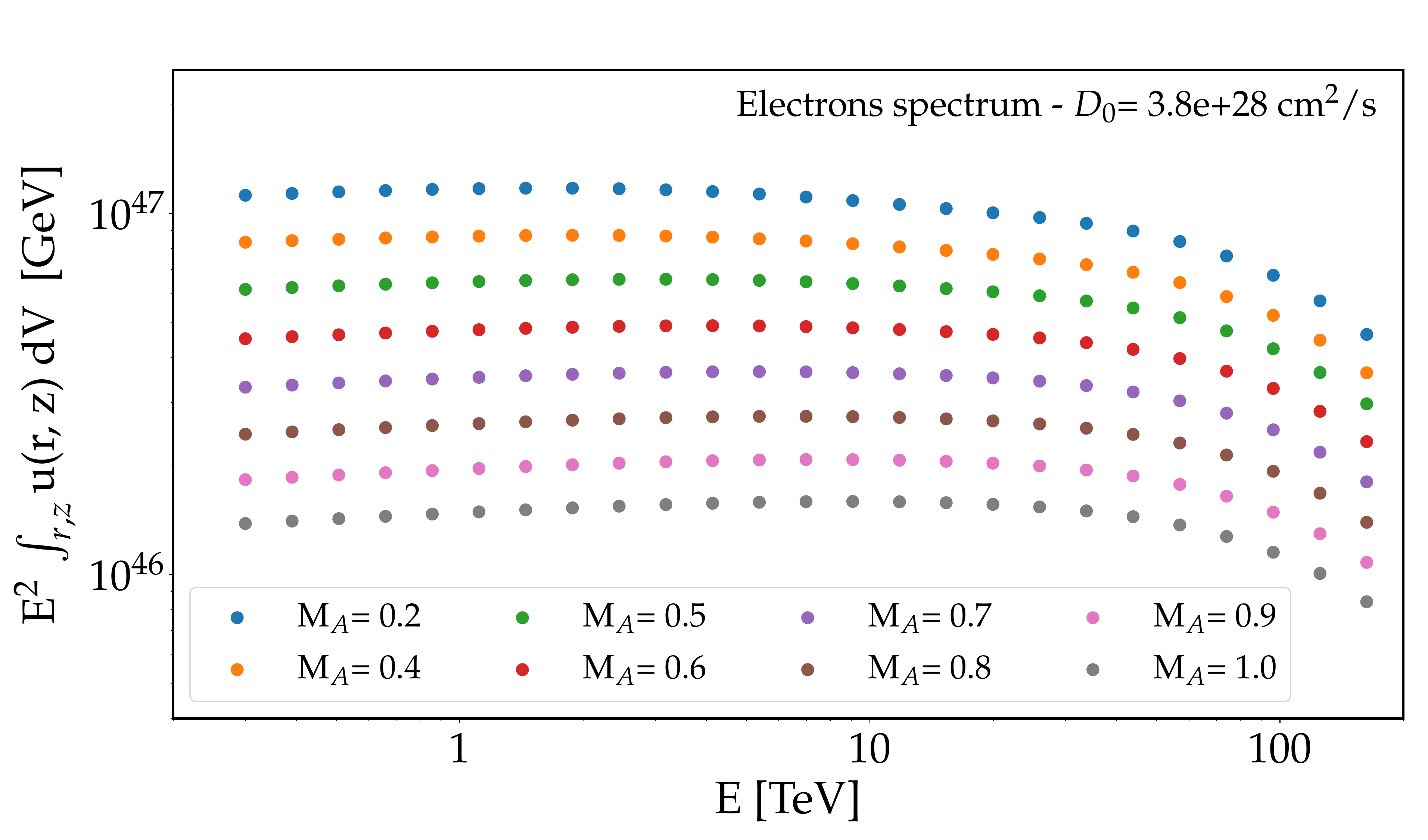}
    \hspace{0.2cm}
    \includegraphics[width=0.48\linewidth, height = 0.287\textwidth]{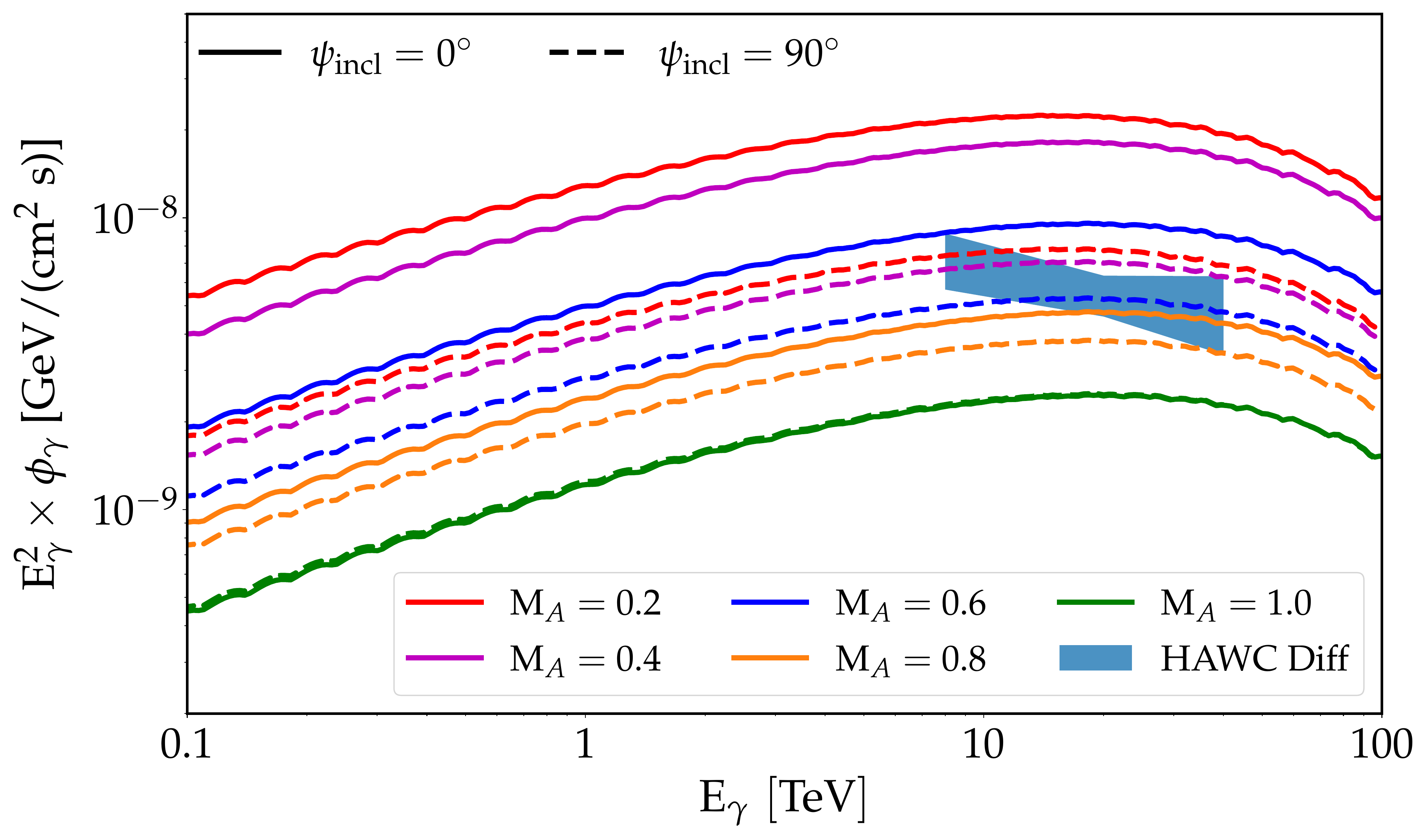}
\caption{\textbf{Left}: Electron spectra for $M_A$ values from $0.1$ to $1$, integrated within $5.5^{\circ}$ around the pulsar. \textbf{Right:} Predicted gamma-ray emissivity spectra for various values of $M_A$  and compared to HAWC data within a window of $5.5^{\circ}$ around the pulsar. In each case the pulsar efficiency is set to 1, which means that models which exceed the HAWC data are potentially compatible with HAWC observations, while models that fall below the HAWC data are in tension with HAWC observations due to energetic considerations.}
\label{fig:e_spectrum}
\end{figure}
Equation~\eqref{eq:eq_numeric} is not directly solvable in its form, so we solve it applying the \textit{alternating-direction implicit} (ADI) method. This requires converting Equation~\eqref{eq:eq_numeric} into two equations (concretely, we follow the Peaceman and Rachford scheme~\cite{Peaceman-Rachford}), which differentiate in steps of $\Delta t/2$ (implying that in the injection $\Delta t \rightarrow \Delta t/2$).
Eq.~\ref{eq:ADI_1} solves the equation implicitly in the $\bm{z}$-direction while explicitly in the $\bm{r}$-direction, and Eq.~\ref{eq:ADI_2} evolves solving the equation explicitly in the $\bm{z}$-direction and implicitly for the $\bm{r}$-direction. The coefficients involved, from A to S (A' to S'), result from rearranging the new equations and keeping all the discretized terms at time $\tau + dt/2$ ($\tau + dt$) on the left-hand side and all the terms at $\tau$ ($\tau + dt/2$) on the right-hand side. These coefficients are commonly referred to as \textit{Crank-Nicolson coefficients}. Finally, each of these equations is easily solvable using (tri-diagonal) matrix operations.

These two equations can be expressed as:
\begin{equation}
\label{eq:ADI_1}
    A u^{\tau+1/2}_{i,k,\epsilon} + B u^{\tau+1/2}_{i,k+1,\epsilon} + C u^{\tau+1/2}_{i,k-1,\epsilon} = D u^{\tau}_{i,k,\epsilon} + E u^{\tau}_{i+1,k,\epsilon} + F u^{\tau}_{i-1,k,\epsilon} + G u^{\tau}_{i,k,\epsilon+1} + \Delta t/2 \cdot S 
\end{equation}

and 
\begin{equation}
\label{eq:ADI_2}
    A' u^{\tau+1}_{i,k,\epsilon} + B' u^{\tau+1}_{i+1,k,\epsilon} + C' u^{\tau+1}_{i-1,k,\epsilon} = D' u^{\tau+1/2}_{i,k,\epsilon} + E' u^{\tau+1/2}_{i,k+1,\epsilon} + F' u^{\tau+1/2}_{i,k-1,\epsilon} + G' u^{\tau+1/2}_{i,k,\epsilon+1} + \Delta t/2 \cdot S ,
\end{equation}
where the terms from $A$ to $G'$ are the \textit{Crank-Nicolson coefficients} associated to each of the density bin indexes.

\begin{figure}[ht]
    \centering
    \includegraphics[width=0.55\linewidth]{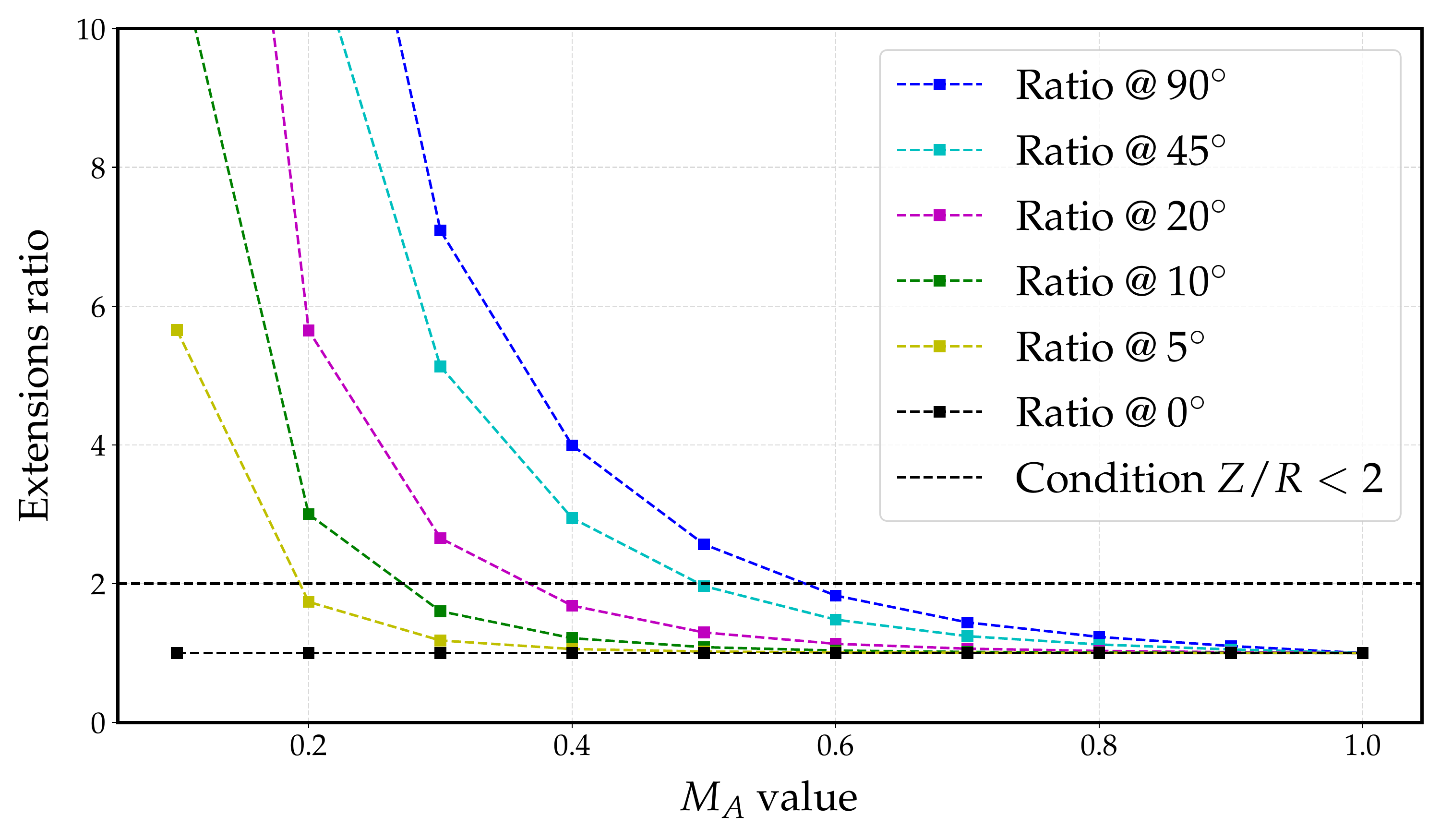}
\caption{The ratio of the extension of the TeV halo in the Z-axis divided by the extension of the halo in the R-axis for different inclination angles ($\psi_{\mathrm{incl}}$). The horizontal dashed line represents the default condition imposed to fulfill TeV halo observations is $Z/R<2$, an upper-limit set in this work.}
\label{fig:s_condition}
\end{figure}

The left panel of Figure~\ref{fig:e_spectrum} shows the propagated spectrum of electrons.
This emission is proportional to the flux of gamma rays generated from IC process, as shown in Figure~\ref{fig:e_spectrum} (right), where we also include the Geminga gamma-ray flux measured by HAWC~\cite{HAWC:2017kbo}. Here we observe that at about $50$ TeV, the emission starts to be suppressed by the Klein-Nishina effect and the injection cut-off. In this figure the efficiency is set to $1$.
In Figure~\ref{fig:s_condition} we show the ratio of the extension of the simulated TeV halo in the projected Z-axis divided by the extension of the halo in the projected R-axis for different inclination angles ($\psi_{\mathrm{incl}}$) with respect to our LoS. As this ratio becomes bigger, bigger is the asymmetry of the object. Here, we see that models with $M_A$ smaller than $\sim \, 0.6$ become increasingly incompatible with the isotropy of TeV halo observations as seen from most inclination angles. 

\begin{figure*}[!th]
    \begin{subfigure}
        \centering
        \includegraphics[width=0.48\linewidth]{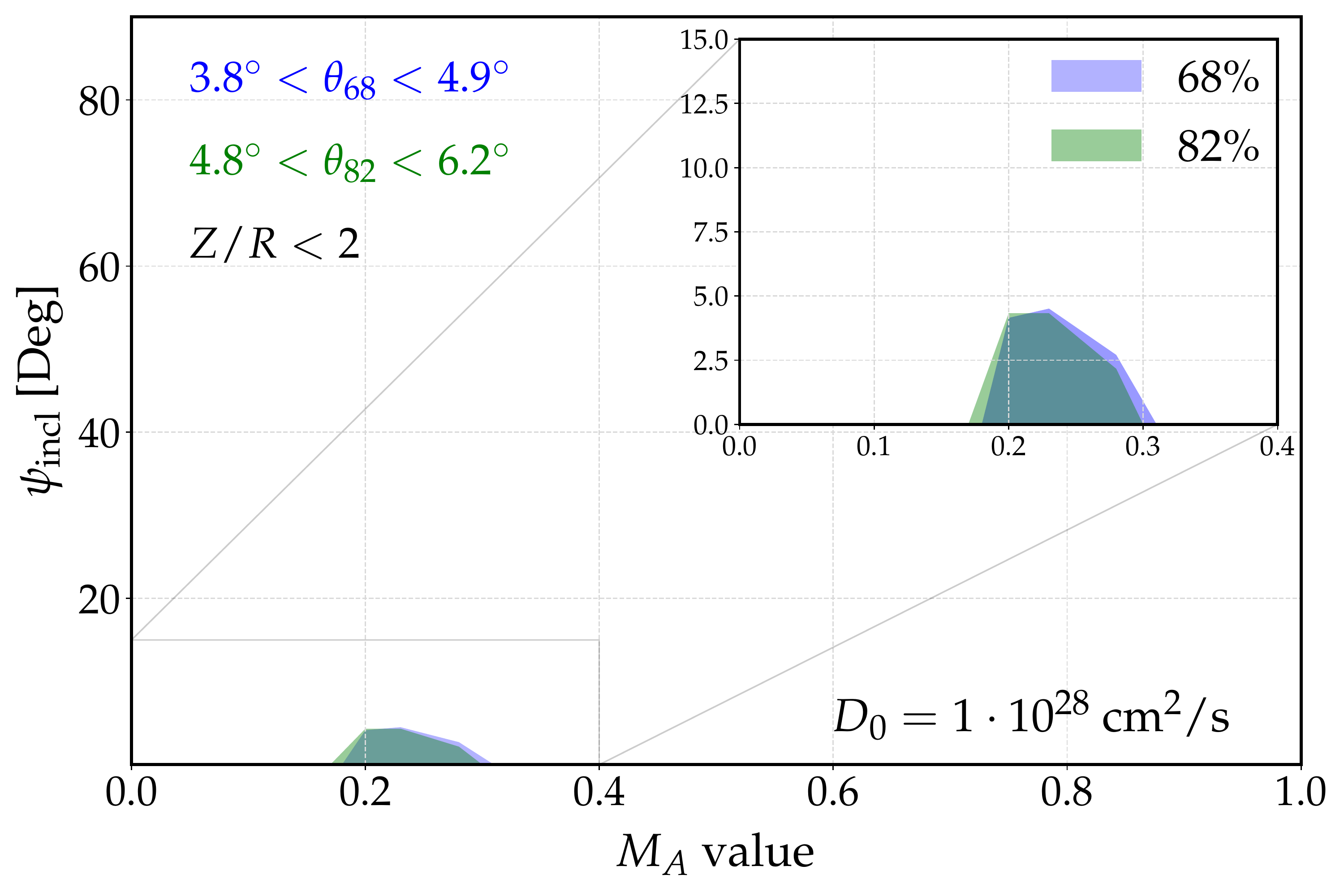}
    \end{subfigure}%
    \begin{subfigure}
        \centering
        \includegraphics[width=0.48\linewidth]{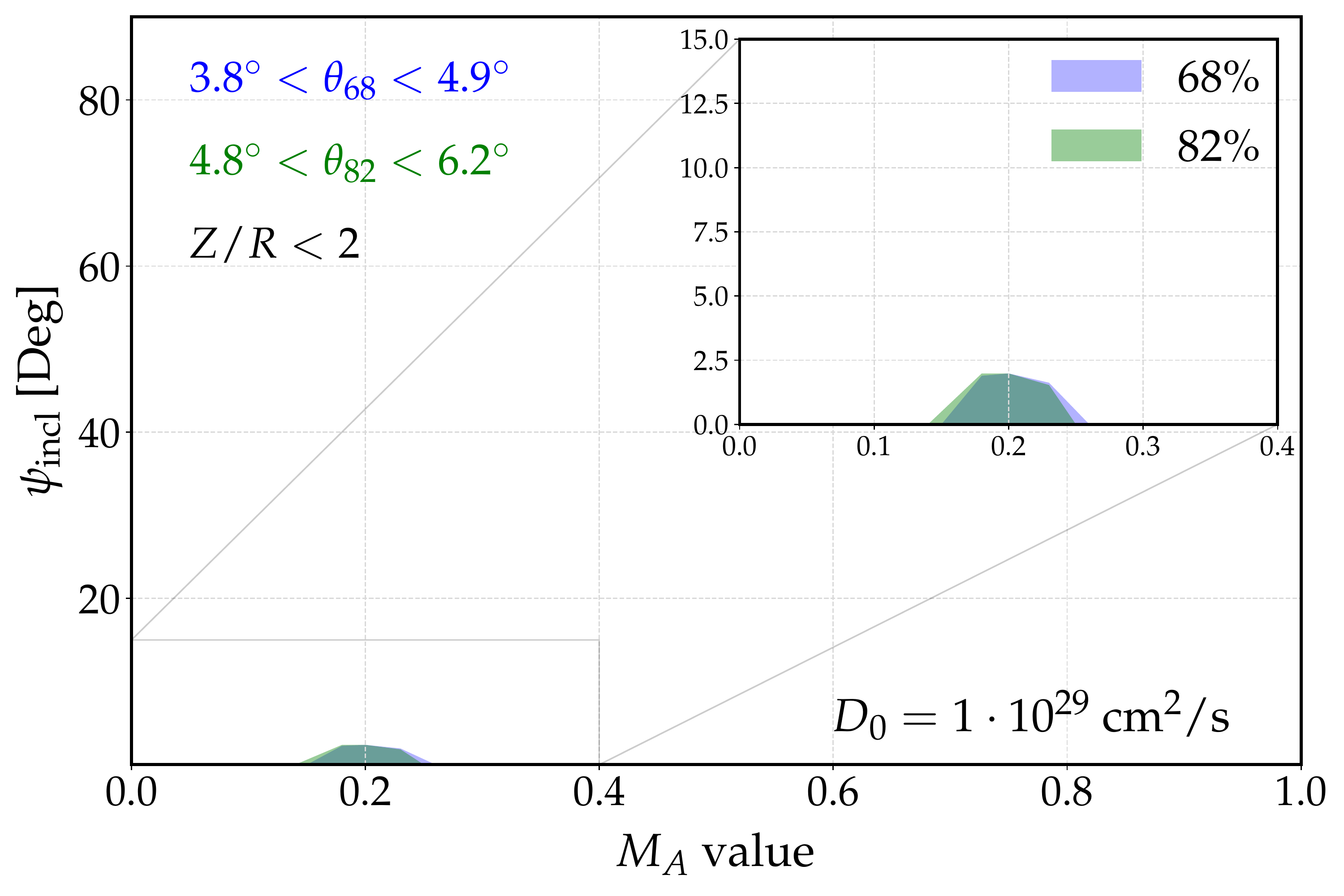}
    \end{subfigure}%
\caption{ \label{fig:D0+-} Analogous to Figure~\ref{fig:Allowed_space}, but for normalization of the diffusion coefficient set to $D_0 = 10^{28} \, \mathrm{cm^{2} \, s^{-1}}$ (left panel) and $D_0 = 10^{29} \, \mathrm{cm^{2} \, s^{-1}}$ (right panel). These models are both extreme, already residing in significant tension with cosmic-ray secondary-to-primary ratios.}
\end{figure*}

In Figure~\ref{fig:D0+-} we show the constraint on the TeV halo population in the parameter space of $M_A$ and $\psi_{\mathrm{incl}}$. Only a very reduced space of inclination angles ($\psi_{\mathrm{incl}} < 5^{\circ}$) is able to simultaneously account for the radial size and measured symmetry of a typical TeV halo. The left panel shows the allowed $M_A$ and $\psi_{\mathrm{incl}}$ parameter space assuming a normalization of the diffusion coefficient $D_0=10^{28} \, \mathrm{cm^2 \, s^{-1}}$ while the right panel assumes a normalization $D_0=10^{29} \, \mathrm{cm^2 \, s^{-1}}$. As discussed in the main text, we observe that even considering slightly different diffusion coefficient normalizations the allowed parameter space able to explain the \textit{size} and \textit{symmetry} conditions imposed is very limited. We also observe here that as we go to lower normalizations, the parameter space increases, meaning that inhibited diffusion is preferred.

\section{Line-of-sight integrated emission - 2D Projected halos}\label{sec:AppC}

\begin{figure*}[!th]
    \centering
    \includegraphics[width=0.47\linewidth]{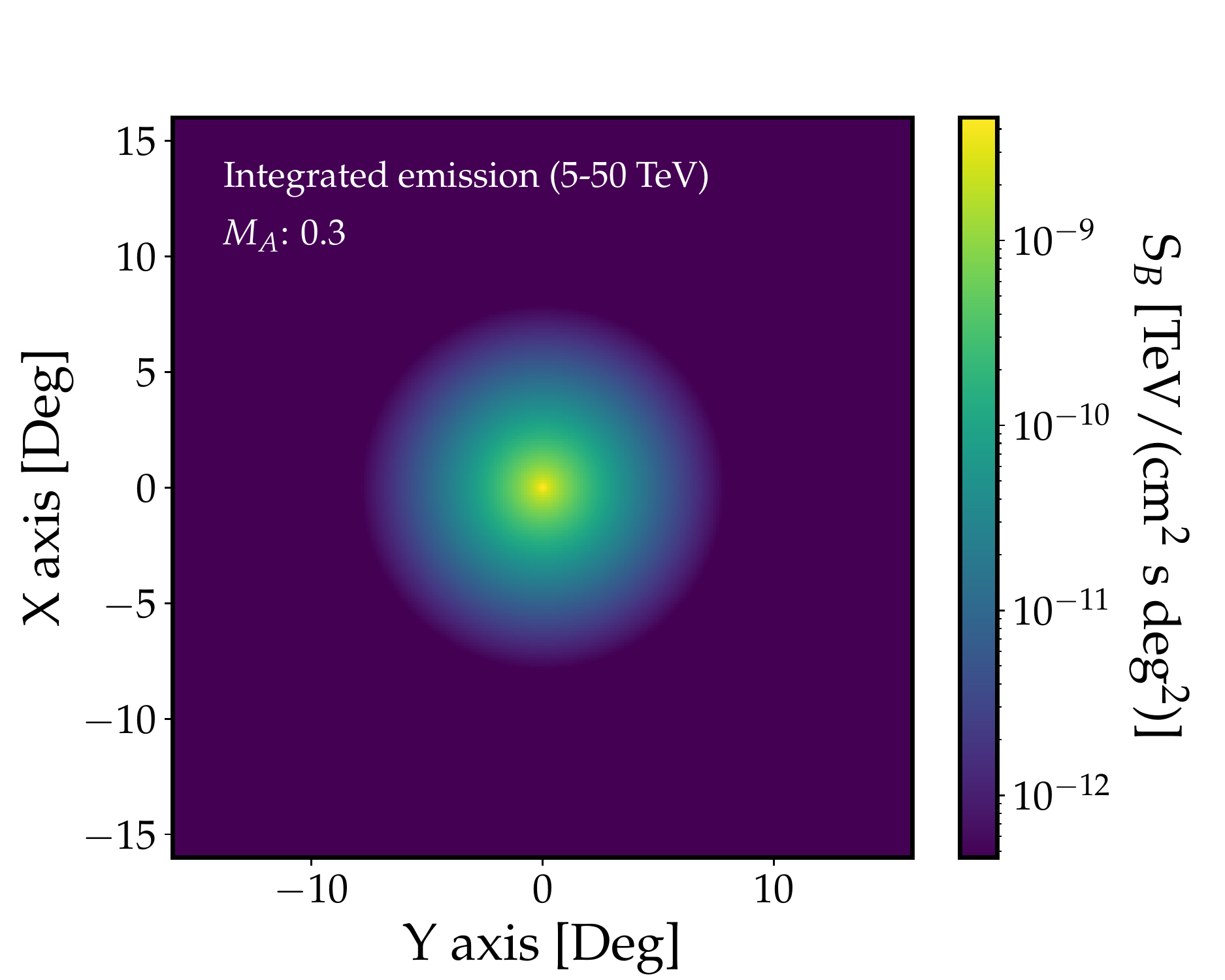}
    \hspace{0.4cm}
    \includegraphics[width=0.47\linewidth]{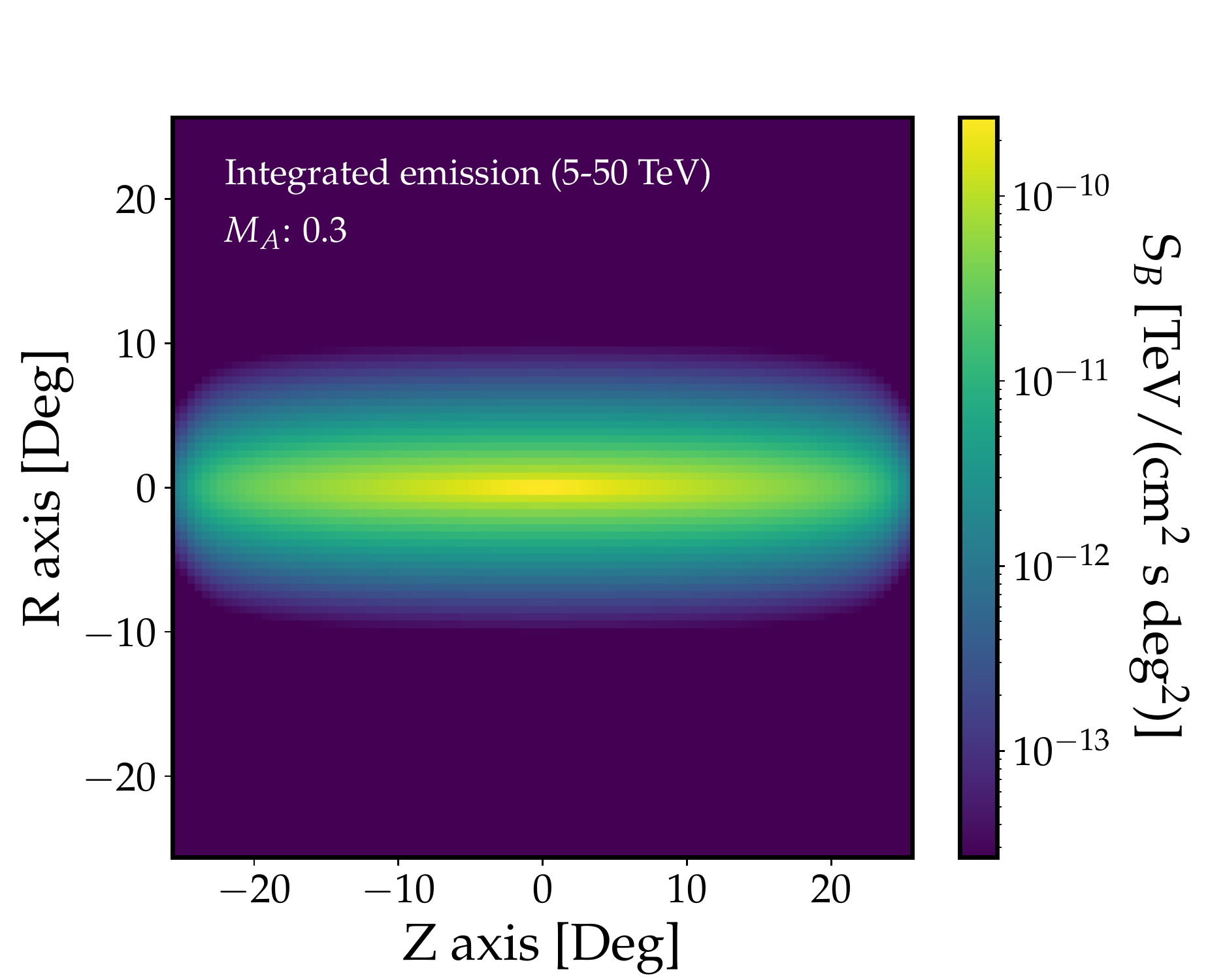}
\caption{2D Projected gamma-ray surface brightness integrated from $5$ to $50$~TeV for the simulated halo in the M$_A=0.3$ case, for inclination angles of $\psi_{\mathrm{incl}}=0^{\circ}$ (left panel) and $\psi_{\mathrm{incl}}=90^{\circ}$ (right panel). In the $\psi_{\mathrm{incl}}=0^{\circ}$ case we denote the projected axes as $X$ and $Y$.} 
\label{fig:2DImages}
\end{figure*}

In this section, we report the 2D projected surface brightness of the simulated halos for the M$_A=0.3$ case, both for $\psi_{\mathrm{incl}}=0^{\circ}$ and $\psi_{\mathrm{incl}}=90^{\circ}$, since these images can be directly compared to the ones reported by experiments (Fig.~\ref{fig:2DImages}).

\newpage
\section{Expected emission profile integrated along the LoS for spherical and non-spherical objects}\label{sec:AppB}

In Figure~\ref{fig:los} (top, left), we notice that in the case $M_A$~=~0.1 and $\psi_{\mathrm{incl}} = 0^\circ$, where we have highly inhibited diffusion in the observable spatial directions, which produces an angular dependence and isotropy similar to observations, we still do not achieve a morphological profile that looks like the $1/r$ profile observed for the Geminga source. In this appendix, we provide an analytic calculation which illustrates why such a phenomenon is expected. 

In order to produce our analytic calculation, we operate under the assumption of a time-independent cosmic-ray injection. While such an injection process is in general not justified, however, we should evaluate it in a window corresponding to the typical size observed for TeV halos, which we here report as $\sim 25 \, \mathrm{pc}$. In such a small window, the perpendicular diffusion timescale can be estimated as $\tau_{\perp, \mathrm{diff}} = \frac{(25 \, \mathrm{pc})^2}{4 \cdot D_{\perp}(E = 100 \, \mathrm{TeV})} \simeq 2 \, \mathrm{kyr}$, where we considered the diffusion coefficient corresponding to $M_A = 0.4$, which reduces the standard diffusion coefficient approximately by two orders of magnitude, as reported by HAWC~\citep{HAWC:2017kbo}. Therefore, we are observing the last $2 \, \mathrm{kyr}$ of the TeV-halo evolution, during which, since $\tau_{\perp, \mathrm{diff}} < \tau_0$, the luminosity does not appreciably change. As a consequence, for our purposes we can effectively consider the emission as being constant over time.

To compute the radial profile, we will consider the Green's function for the transport equation for protons, with no losses involved, which is again justified by the very fast diffusion of high energy particles through our $60$~pc region. This solution represents the CR distribution resulting from a point-like source $\mathcal{S}$ injecting particles in a time burst, $\mathcal{S}(\bm{r}, t, E) = \delta(\bm{r})\delta(t)Q(E)$ --- and we will integrate it over time. We will finally see that introducing the losses for leptons does not change the expected result. The usual Green's function for protons propagated in $n$-dimensions is a Gaussian spreading in space, and its integral over time reads~\citep{1964ocr..book.....G}:
\begin{equation}
    w_n(r, E) = \int_0^{\infty} dt \, \frac{e^{-r^2 \big/ 4D(E)t}}{\left(4 \pi D(E) t \right)^{n/2}}
\end{equation}
where $n$ is the dimensionality of the problem and $r^2 = \sqrt{x^2_1 + ... + x^2_n}$ and from now on we will drop the dependence on $E$, as it is not relevant for the present calculation.

With the following change of variables:
\begin{equation*}
    \frac{r^2}{4Dt} \equiv x \qquad \longrightarrow \qquad 
    \begin{aligned}
        &\frac{dx}{dt} = \frac{r^2}{4D} \times \left( -\frac{1}{t^2} \right) \\
        &\Rightarrow \; dt = -\frac{t}{x} \, dx = -\frac{1}{x} \frac{r^2}{4Dx}\, dx
    \end{aligned}
\end{equation*}
we immediately obtain:
\begin{equation}
\begin{aligned}
    w_n(r, E) &= - \int_{\infty}^0 dx \, e^{-x} \, \frac{x^{n/2}}{(\pi r^2)^{n/2}} \times \frac{r^2}{4 D x^2} = \int^{\infty}_0 dx \, e^{-x} \, \frac{x^{n/2}}{(\pi r^2)^{n/2}} \times \frac{r^2}{4 D x^2} \\
    &= \frac{1}{4 D \, \pi^{n/2} \, r^{n-2}} \times \int_0^{\infty} dx \, e^{-x} \times x^{(n-4)/2} \\
    &= \frac{1}{4 D \, \pi^{n/2} \, r^{n-2}} \times \int_0^{\infty} dx \, e^{-x} \times x^{(n-2)/2 - 1}  \equiv \frac{1}{4 D \, \pi^{n/2} \, r^{n-2}} \times \Gamma\left( \frac{n-2}{2} \right),
\end{aligned}
\end{equation}
we slightly rearranged the integral in order to match the definition of the Euler's Gamma function, \mbox{$\Gamma(z) = \int_0^{\infty} dx \, e^{-x} \times x^{z-1}$}. As a consequence, we immediately find:
\begin{equation}\label{eq:app_CR_densities}
    \begin{aligned}
        &n=3 \qquad \longrightarrow \qquad w_3(r, E) = \frac{1}{4 \pi^{3/2} D r} \times \Gamma\left( \frac{1}{2} \right) = \frac{1}{4 \pi D r} \\
        &n=2 \qquad \longrightarrow \qquad w_2(r, E) = \frac{1}{4 \pi D} \times \Gamma(0) = \mathrm{undefined} \\
        &n=1 \qquad \longrightarrow \qquad w_1(r, E) = \frac{1}{4 \pi ^{1/2} D r^{-1/2}} \times \Gamma \left( - \frac{1}{2} \right) \rightarrow +\infty.
    \end{aligned}
\end{equation}

This implies that, in general, we can expect a $\sim 1/r$ radial profile only from a point-like source constantly injecting with spherical symmetry. The physical interpretation of the previous result can be achieved as the number of times that a particle revisits the source location in a $n$D phase-space. In the physical finite-time simulation that we are performing, this implies that we are not able to give a mathematical expectation for the radial profile.

If we added a loss term in the transport equation, an additional complementary error function, $\mathrm{erfc}(x) = \frac{2}{\sqrt{\pi}} \int_x^{\infty}dt \, e^{-t^2}$, would appear multiplied by the Green's function~\citep{PhysRevD.52.3265}, such that:
\begin{equation}
    w_{3, \mathrm{leptons}}(r, E) = \frac{1}{4 \pi D r}\times \mathrm{erfc} \left( \frac{r}{\sqrt{4 D \times t_{\mathrm{loss}}}} \right),
\end{equation}
where $t_{\mathrm{loss}}$ is the time needed for our cosmic-ray leptons to loose all their energy after being released.

However, for $E_e = 10 \, (100) \, \mathrm{TeV}$ leptons, in a few $\mu \mathrm{G}$ magnetic field, we have roughly $t_{\mathrm{loss}} \simeq 2.5 \times 10^4 \, (10^3) \, \mathrm{yr}$, namely they can propagate for $\sqrt{4 D \times t_{\mathrm{loss}}} \sim 490 \, (219) \, \mathrm{pc}$, with $D$ calculated as the standard ISM-value used in this work. As a consequence, $r_{\mathrm{max}} = 60 \, \mathrm{pc} \ll \sqrt{4 D \times t_{\mathrm{loss}}}$ for the energies under study and $\mathrm{erfc} (x \ll 1) \sim \mathcal{O}(1)$, which makes us recover the CR density $w_3 (r, E)$ derived in Equation \eqref{eq:app_CR_densities}.

\newpage

\end{document}